\documentclass[journal,11pt,draft,onecolumn]{IEEEtran}
\usepackage{subfig}
\usepackage{fullpage}
\usepackage{amssymb}
\usepackage{cite}
\usepackage[pdftex]{graphicx}
\graphicspath{{./}{figures/}}
\usepackage[table]{xcolor}
\usepackage{amsmath}
\usepackage{amssymb}
\usepackage{url}
\usepackage{setspace}
\usepackage{fixltx2e}
\usepackage{floatrow}
\usepackage[font=small,labelfont=bf]{caption}
\newfloatcommand{capbtabbox}{table}[][\FBwidth]

\def\cat{{\mathbf{C}}}
\usepackage{blindtext}
\def\eg{\emph{e.g. }}
\def\ie{\emph{i.e. }}
\def\etal{\emph{et al. }}

\def\mod{\textrm{ mod }}
\newcommand{\detm}[1]{\left|{#1}\right|}

\usepackage{algorithm}
\usepackage{algorithmic}

\begin{document}
\title{A New Family of Generalized 3D Cat Maps}
\author{Yue~Wu,~\IEEEmembership{Student Member,~IEEE,}
%        Yicong~Zhou,~\IEEEmembership{Member,~IEEE,}
        {Sos~Agaian},~\IEEEmembership{Senior Member,~IEEE,}\\
        and Joseph~P.~Noonan,~\IEEEmembership{Life Member,~IEEE,}
\thanks{Yue Wu and Joseph P. Noonan is with the Department of Electrical and Computer Engineering, Tufts University, Medford, MA 02155, United States; e-mail: ywu03@ece.tufts.edu.}
%\thanks{Yicong Zhou is with the Department of Computer and Information Science, University of Macau, Macau, China; email: yicongzhou@umac.mo}
\thanks{Sos Agaian is with the Electrical and Computer Engineering, University of Texas at San Antonio, San Antonio, TX 78249, United States.}
}

%\markboth{A Paper Draft Submitted to IEEE Transactions on Multimedia}{Wu \etal}
\maketitle

\begin{abstract}
Since the 1990s chaotic cat maps are widely used in data encryption, for their very complicated dynamics within a simple model and desired characteristics related to requirements of cryptography. The number of cat map parameters and the map period length after discretization are two major concerns in many applications for security reasons. In this paper, we propose a new family of 36 distinctive 3D cat maps with different spatial configurations taking existing 3D cat maps \cite{CatLian,3DCat,CatLiu,CatPan} as special cases. Our analysis and comparisons show that this new 3D cat maps family has more independent map parameters and much longer averaged period lengths than existing 3D cat maps. The presented cat map family can be extended to higher dimensional cases.
\end{abstract}

\IEEEpeerreviewmaketitle

\section{Introduction}
In the last decade, many efforts have been recognized to study behaviors of dynamic systems and related applications. As one type of dynamic systems with very complicated behaviors, chaos systems are widely reported in mathematics, physics, engineering, economics, and biology.
Especially for cryptography and encryption, chaotic cryptosystems \cite{amigo2007theory} are demonstrated to have many analogous to conventional encryption methods \cite{3DCat}, \eg a conventional algorithm is sensitive to keys, while a chaotic system is sensitive to its initial values and parameters. Chaotic encryption systems also provide better solutions for image encryption where the conventional method are not suitable due to many intrinsic features of images \cite{3DCat}, \eg bulk data nature and high correlations between pixels. Consequently, numerous chaotic encryption algorithms have been proposed in literature based on various principles\cite{3DCat,Pareek2006,kumar2010substitution,Kwok2007}. Among these applications, Arnold's cat map and its variants are commonly employed as a fundamental building block for data encryption \cite{CatLian,3DCat,CatLiu,CatPan,kumar2010substitution,Fu2011,kanso2011novel,liu2011double}, watermarking \cite{5116721}, and pattern recognition \cite{deng2011color}.

Arnold's cat map \cite{3DCat,ford1991arnol} is a discrete chaotic map named after \textit{Vlamimir Arnold}.
% for his contribution of demonstrating the map effect using an image of a cat.
Specifically, it is erogdic and mixing, a C-system, a K-system and a Bernoulli system \cite{ford1991arnol}. Mathematically, Arnold's cat map is defined in Eq. \eqref{eq.CatMap}
\begin{equation}\label{eq.CatMap}
    {x_{n+1}\brack y_{n+1}} = \mathbf{C}{x_{n}\brack y_{n}} \mod 1
\end{equation}
where the transform matrix $\mathbf{C}$ is defined as
\begin{equation}\label{eq.CatMap2D}
    \mathbf{C}= \begin{bmatrix}2&1\\1&1\end{bmatrix}
\end{equation}

In order to achieve higher map randomness and thus a more secure cryptosystem, the cat matrix $\cat$ of Arnold's cat map is commonly replaced by Eq. \eqref{eq.CatMap2P}, where new parameters help increase key spaces to resist brute-force attacks and longer averaged period lengths help improve map randomness to resist statistical attacks.
\begin{equation}\label{eq.CatMap2P}
    \mathbf{C}{^{2D}_{para}}= \begin{bmatrix}1&a\\b&ab+1\end{bmatrix}
\end{equation}
Based on Eq. \eqref{eq.CatMap2P}, many efforts \cite{CatLian,3DCat,CatLiu,CatPan} are put on designing 3D cat maps to further increase the parameter space and map randomness. However, we notice that 1) they failed to consider all possible spatial configurations of a 3D cat map; and 2) their 3D cat map matrix elements are designed in more correlate rather than independent to each other.

In this paper, we propose a new family of 3D cat maps with improvements on existing 3D cat maps \cite{CatLian,3DCat,CatLiu,CatPan}. This new family contains more parameters with longer averaged period lengths and taking these commonly used 3D cat maps as special cases. The rest of the paper is organized as follows: Section II reviews four existing 3D cat maps; Section III introduces our algorithm for generating the new 3D cat map family; Section IV analyzes the averaged period lengths of the proposed 3D cat maps with other 3D cat maps; and Section V concludes the letter.
\section{Existing 3D Cat Maps}
Based on the parametric 2D cat map in Eq. \eqref{eq.CatMap2P}, Lian \etal \cite{CatLian} proposed a parametric 3D cat map $\mathbf{C}{_{L}^{3D}}$ in 2003 by extending 2D cat map in $zx$ plane and $yz$ plane.
\begin{equation}\label{eq.CatMapLian3}
        \mathbf{C}{_{L}^{3D}}= \begin{bmatrix}1&0&a\\bc&1&abc+c\\bcd+b&d&abcd+ab+cd+1\end{bmatrix}
\end{equation}

Later on, in 2004 Chen \etal \cite{3DCat} proposed a family of parametric 3D cat maps defined in Eq. \eqref{eq.CatMapChen3} by composing three fundamental parametric 3D cat maps on $xy$ and $yz$ and $zx$ planes defined in Eqs. \eqref{eq.CatMapChenX}, \eqref{eq.CatMapChenY} and \eqref{eq.CatMapChenZ}, respectively.
\begin{eqnarray}\label{eq.CatMapChen3}
    \mathbf{C}{_{C}^{3D}} = \mathbf{C}{_{C}^{xy}}\mathbf{C}{_{C}^{yz}}\mathbf{C}{_{C}^{zx}} %\\
    %\begin{bmatrix}a_xa_zb_y+1&a_z&a_y+a_xa_z(a_yb_y+1)\\b_z+a_xb_y(a_zb_z+1)&a_zb_z+1&a_yb_z+a_x(a_yb_y+1)(a_zb_z+1)\\b_y(a_xb_x+1)&b_x&(a_xb_x+1)(a_yb_y+1)\end{bmatrix}
\end{eqnarray}
\begin{equation}\label{eq.CatMapChenX}
        \mathbf{C}{_{C}^{xy}}= \begin{bmatrix}1&0&0\\0&1&a_x\\0&b_x&a_xb_x+1\end{bmatrix}
\end{equation}
\begin{equation}\label{eq.CatMapChenY}
        \mathbf{C}{_{C}^{yz}}= \begin{bmatrix}1&0&a_y\\0&1&0\\b_y&0&a_yb_y+1\end{bmatrix}
\end{equation}
\begin{equation}\label{eq.CatMapChenZ}
        \mathbf{C}{_{C}^{zx}} = \begin{bmatrix}1&a_z&0\\b_z&a_zb_z+1&0\\0&0&1\end{bmatrix}
\end{equation}

Based on Chen's cat map, Liu \etal \cite{CatLiu} proposed an improved 3D cat map $\mathbf{C}{_{U}^{3D}}$ in 2008 by introducing new parameters $c$ and $d$ as shown in Eq. \eqref{eq.CatMapLiu3}
\begin{equation}\label{eq.CatMapLiu3}
    \mathbf{C}{_{U}^{3D}} = \begin{bmatrix}1&a&0\\b&ab+1&0\\c&d&1\end{bmatrix}
\end{equation}

Recently in \cite{CatPan}, Pan \etal introduced their 3D cat map $\mathbf{C}{_{P}^{3D}}$ based on Chen's method, but in a different formulation from Liu \etal's.
\begin{equation}\label{eq.CatMapPan3}
    \mathbf{C}{_{P}^{3D}} = \begin{bmatrix}1&a&c\\b&ab+1&bc\\d&abcd&cd+1\end{bmatrix}
\end{equation}
It is worthwhile note that all these cat maps satisfy the condition that the determinant of 3D cat matrix is 1, \ie $$\detm{{\mathbf{C}}^{3D}_{L}}= \detm{{\mathbf{C}}^{3D}_{C}} = \detm{{\mathbf{C}}^{3D}_{U}} = \detm{{\mathbf{C}}^{3D}_{P}} = 1$$
\section{A New Family of 3D Cat Maps}
One common feature of existing 3D cat maps \cite{CatLian,3DCat,CatLiu,CatPan} that they are all extended from the 2D cat map $\cat^{2D}_{para}$ defined in Eq. \eqref{eq.CatMap2P}, which is indeed a general form of 2D cat but only one out of four possible spatial configurations as shown in Eqs. \eqref{eq.CatMap2D1}-\eqref{eq.CatMap2D4}:
\begin{equation}\label{eq.CatMap2D1}
    \cat{^{2D}_1} = \cat{^{2D}_{para}}%\begin{bmatrix}1&a\\b&ab+1\end{bmatrix}
\end{equation}
\begin{equation}\label{eq.CatMap2D2}
    \cat{^{2D}_2}  = \begin{bmatrix}a&1\\ab-1&b\end{bmatrix}
\end{equation}
\begin{equation}\label{eq.CatMap2D3}
    \cat{^{2D}_3}  = \begin{bmatrix}a&ab-1\\1&b\end{bmatrix}
\end{equation}
\begin{equation}\label{eq.CatMap2D4}
    \cat{^{2D}_4} = \begin{bmatrix}ab+1&a\\b&1\end{bmatrix}
\end{equation}
As can be seen, these configurations are inequivalent to each other in general because none of them has the unitary element located in the same position. However, spatial configurations are fail to be considered in previous 3D cat maps \cite{CatLian,3DCat,CatLiu,CatPan}.

Meanwhile, we also notice that constructing 3D cat maps by directly extending the 2D cat map $\cat^{2D}_{para}$ as done in \cite{CatLian,3DCat,CatLiu,CatPan} is not necessary, because all what we need is to construct a $3\times 3$ matrix $\cat^{3D}$ with the constraint that its determinant is 1 \cite{3DCat}. In other words, we are looking for a $3\times 3$ matrix $\cat^{3D}_W$ with the symbol set $\mathfrak{S} = \{\mathfrak{a},\mathfrak{b},\mathfrak{c},\mathfrak{d},\mathfrak{e},\mathfrak{f},\mathfrak{g},\mathfrak{h},\mathfrak{i}\}$ as shown in Eq. \eqref{eq.matrix3} whose determinant satisfies Eq. \eqref{eq.det}. \begin{equation}\label{eq.matrix3}
    {\cat^{3D}_W} = \begin{bmatrix}\mathfrak{a}&\mathfrak{b}&\mathfrak{c}\\\mathfrak{d}&\mathfrak{e}&\mathfrak{f}\\\mathfrak{g}&\mathfrak{h}&\mathfrak{i}\end{bmatrix}
\end{equation}
\begin{equation}\label{eq.det}
\detm{{\cat^{3D}_W}} = \mathfrak{a}\mathfrak{e}\mathfrak{i}+\mathfrak{b}\mathfrak{f}\mathfrak{g}+\mathfrak{c}\mathfrak{d}\mathfrak{h}-\mathfrak{c}\mathfrak{e}\mathfrak{g}-\mathfrak{b}\mathfrak{d}\mathfrak{i}-\mathfrak{a}\mathfrak{f}\mathfrak{h}=1
\end{equation}

We describe a general solution to the above constraint problem in Algorithm 1. This algorithm is able to generate 36 spatial configurations specified by parameter $u$ and $m$ for 3D cat maps, each configuration with six other independent parameters controlling cat matrix elements.

\begin{algorithm}
\small
\begin{algorithmic}
\REQUIRE $u$ is an integer in $\{1,2,\cdots,9\}$
\REQUIRE $m$ is an integer in $\{1,2,3,4\}$
\ENSURE $\mathbf{C}{^{3D}_{W_{um}}}$ is a 3D Cat matrix with $\detm{\cat^{3D}_{W_{um}}} = 1$
\STATE 1. Set symbol $\mathfrak{S}_u$ to 1 and compute the target determinant $ \mathfrak{a}\mathfrak{e}\mathfrak{i}+\mathfrak{b}\mathfrak{f}\mathfrak{g}+\mathfrak{c}\mathfrak{d}\mathfrak{h}-\mathfrak{c}\mathfrak{e}\mathfrak{g}-\mathfrak{b}\mathfrak{d}\mathfrak{i}-\mathfrak{a}\mathfrak{f}\mathfrak{h} = 1$ by substituting $\mathfrak{S}_u = 1$
\STATE 2. Form the four symbols in the positive or negative diagonal containing the unitary symbol $\mathfrak{S}_u$ to set $\mathfrak{F}$
\STATE 3. Collect the terms containing the symbol $\mathfrak{F}_m$ on the left side of the equation and leave all the other terms on the right side with a name $rightside$
\STATE 4. Set the coefficient term of symbol $\mathfrak{F}_m$ to 1 and result in symbol $\mathfrak{F}_m = rightside$.
\STATE 5. Output $\mathbf{C}{^{3D}_{W_{um}}}$ as a 3D cat map.
\end{algorithmic}
%\caption{\textbf{Parametric 3D Cat Matrix \\$\cat^{3D}_{W_{um}} = P3DCM(u,m)$}}
\caption{\textbf{Cat Matrix $\cat^{3D}_{W_{um}}$ Generator}}
\end{algorithm}
\normalsize
%This is a very high-level description of our family of 3D cat maps with only spatial configuration parameters.
Details of why this algorithm works are illustrated by the following example. Assume $u=m=1$, then we first set the unitary symbol $\mathfrak{S}_u=\mathfrak{S}_1=\mathfrak{a} = 1$, \ie we have $$\mathbf{C}{^{3D}_{W_{11}}} =  \begin{bmatrix}1&\mathfrak{b}&\mathfrak{c}\\\mathfrak{d}&\mathfrak{e}&\mathfrak{f}\\\mathfrak{g}&\mathfrak{h}&\mathfrak{i}\end{bmatrix}$$.
Consequently, the target determinant equation $\detm{\mathbf{C}{^{3D}_{W_{11}}}} = 1$ become $$\mathfrak{e}\mathfrak{i}+\mathfrak{b}\mathfrak{f}\mathfrak{g}+\mathfrak{c}\mathfrak{d}\mathfrak{h}-\mathfrak{c}\mathfrak{e}\mathfrak{g}-\mathfrak{b}\mathfrak{d}\mathfrak{i}-\mathfrak{f}\mathfrak{h} = 1$$
We then obtain a symbol set $\mathfrak{F} = \{\mathfrak{e},\mathfrak{i},\mathfrak{f},\mathfrak{h}\}$ for those either along the positive or the negative diagonal containing the symbol $\mathfrak{S}_1 $; and collect all terms containing symbol $\mathfrak{F}_m=\mathfrak{F}_1 = \mathfrak{e}$ on the left side of the equation, leaving all the other terms on the right side, namely
%\begin{equation}\label{eq.example}
 $$ (\mathfrak{i}-\mathfrak{c}\mathfrak{g})\mathfrak{e} = 1-\mathfrak{b}\mathfrak{f}\mathfrak{g}-\mathfrak{c}\mathfrak{d}\mathfrak{h}+\mathfrak{b}\mathfrak{d}\mathfrak{i}+\mathfrak{f}\mathfrak{h}$$
%\end{equation}
Next we set the coefficient term of symbol $\mathfrak{e}$ to 1, \ie $\mathfrak{i} = \mathfrak{c}\mathfrak{g}+1$, and the simplified equation implies that $$\mathfrak{e} = 1-\mathfrak{b}\mathfrak{f}\mathfrak{g}-\mathfrak{c}\mathfrak{d}\mathfrak{h}+\mathfrak{b}\mathfrak{d}(\mathfrak{c}\mathfrak{g}+1)+\mathfrak{f}\mathfrak{h}$$
And we finish the 3D cat map construction, because $\mathbf{C}{^{3D}_{W_{11}}}$ in Eq. \eqref{eq.W11} is already of a $3\times 3$ matrix with determinant 1.
\begin{equation}\label{eq.W11}
\mathbf{C}{^{3D}_{W_{11}}} = \begin{bmatrix}1&\mathfrak{b}&\mathfrak{c}\\\mathfrak{d}&\mathfrak{b}\mathfrak{d}+\mathfrak{f}\mathfrak{h}-\mathfrak{b}\mathfrak{f}\mathfrak{g}-\mathfrak{c}\mathfrak{d}\mathfrak{h}+\mathfrak{b}\mathfrak{c}\mathfrak{d}\mathfrak{g}+1&\mathfrak{f}\\\mathfrak{g}&\mathfrak{h}&\mathfrak{c}\mathfrak{g}+1\end{bmatrix}\end{equation}
\normalsize

The other three variants with unitary element $a = 1$ of the 3D cat map family are shown below:
\begin{equation}\label{eq.W12}
\mathbf{C}{^{3D}_{W_{12}}} = \begin{bmatrix}1&\mathfrak{b}&\mathfrak{c}\\\mathfrak{d}&\mathfrak{b}\mathfrak{d}+1&\mathfrak{f}\\\mathfrak{g}&\mathfrak{h}&\mathfrak{c}\mathfrak{g}+\mathfrak{f}\mathfrak{h}-\mathfrak{b}\mathfrak{f}\mathfrak{g}-\mathfrak{c}\mathfrak{d}\mathfrak{h}+\mathfrak{b}\mathfrak{c}\mathfrak{d}\mathfrak{g}+1\end{bmatrix}\end{equation}
\begin{equation}\label{eq.W13}
\mathbf{C}{^{3D}_{W_{13}}} = \begin{bmatrix}1&\mathfrak{b}&\mathfrak{c}\\\mathfrak{d}&\mathfrak{e}&\mathfrak{c}\mathfrak{d}+1\\\mathfrak{g}&\mathfrak{b}\mathfrak{g}+\mathfrak{e}\mathfrak{i}-\mathfrak{c}\mathfrak{e}\mathfrak{g}-\mathfrak{b}\mathfrak{d}\mathfrak{i}+\mathfrak{b}\mathfrak{c}\mathfrak{d}\mathfrak{g}-1&\mathfrak{i}\end{bmatrix}\end{equation}
\begin{equation}\label{eq.W14}
\mathbf{C}{^{3D}_{W_{14}}} = \begin{bmatrix}1&\mathfrak{b}&\mathfrak{c}\\\mathfrak{d}&\mathfrak{e}&\mathfrak{c}\mathfrak{d}+\mathfrak{e}\mathfrak{i}-\mathfrak{c}\mathfrak{e}\mathfrak{g}-\mathfrak{b}\mathfrak{d}\mathfrak{i}+\mathfrak{b}\mathfrak{c}\mathfrak{d}\mathfrak{g}-1\\\mathfrak{g}&\mathfrak{b}\mathfrak{g}+1&\mathfrak{i}\end{bmatrix}\end{equation}
\normalsize
In general, a symbolic 3D cat map of the possible 36 spatial configurations can be easily obtained in a similar manner by feeding different $u$s and $m$s in Algorithm 1, each configuration contains six independent parameters like those in Eqs. \eqref{eq.W11}-\eqref{eq.W14}. Therefore, each new parametric 3D cat map is associated with eight independent parameters, two controlling spatial configurations and six controlling matrix elements. Consequently, this new family of 3D cat maps have more parameters but less correlated matrix elements than existing 3D cat maps \cite{CatLian,3DCat,CatLiu,CatPan}. Detailed comparisons about parameters and matrix elements between different 3D cat maps are summarized in Table \ref{tab.independence}.
\begin{table}[h]
\small
\centering
\caption{Elements and Parameters in 3D Cat Maps}\label{tab.independence}
\begin{tabular}{r|lllll}
\hline\hline
& \multicolumn{5}{c}{\textbf{Existing 3D Cat Maps}}\\
\textbf{Number of Items} & ${\cat^{3D}_L}$& ${\cat^{3D}_{C}}$& ${\cat^{3D}_U}$& ${\cat^{3D}_P}$ & ${\cat^{3D}_W}$\\\hline
\textbf{Constant 0 Elements} & 1  & 0 & 2 & 0 & 0\\
\textbf{Constant 1 Elements} & 2  & 0 & 2 & 1 & 1\\
\textbf{1 Parameter Elements} & 2  & 2 & 4 & 4 & 6\\
\textbf{2+ Parameter Elements} & 4  & 7 & 1 & 4 & 2\\\hline
\textbf{Parameters} & 4 & 6 & 4 & 4 & 8\\
\textbf{Spatial Configurations} & 1 &1 &1 &1 &36\\ \hline\hline
\end{tabular}
\end{table}

It is worthwhile to note that the proposed 3D cat map family includes previous 3D cat maps \cite{CatLian,3DCat,CatLiu,CatPan} as special cases as shown in Table \ref{tab.correspondence}, where each existing 3D cat map $\cat^{3D}$ can be denoted by $\cat^{3D}_{W_{um}}$ with the eight parameters listed the table and symbol $\star$ indicates either constant elements or dependent elements on two or more parameters. For example, $\cat^{3D}_P$ is a special case of $\cat^{3D}_{W_{um}}$ because Eq. \eqref{eq.Eexample} holds.
 \begin{equation}\label{eq.Eexample}
    \cat^{3D}_P = \cat^{3D}_{W_{um}} |^{u=1,m=2}_{\mathfrak{b} = a,\mathfrak{c} = c, \mathfrak{d} = b,\mathfrak{f} = bc, \mathfrak{g} = d, \mathfrak{h} = abcd}
\end{equation}
For verification, simply substitute these parameter values in $\cat^{3D}_{W_{12}} $. And we obtain $\star$ elements $\mathfrak{a} = 1$, $\mathfrak{e} = \mathfrak{b}\mathfrak{d}+1 = ab+1$, and $\mathfrak{i} = \mathfrak{c}\mathfrak{g}+\mathfrak{f}\mathfrak{h}-\mathfrak{b}\mathfrak{f}\mathfrak{g}-\mathfrak{c}\mathfrak{d}\mathfrak{h}+\mathfrak{b}\mathfrak{c}\mathfrak{d}\mathfrak{g}+1 = cd+(bc)(abcd)-abcd-cb(abcd)+acbd+1 = cd+1$, which are indeed the three corresponding elements of $\cat^{3D}_P$ defined in Eq. \eqref{eq.CatMapPan3}.
 
%$\cat^{3D}_{W_{12}}$ because with the setting that $\mathfrak{b} = a; \mathfrak{c} = c; \mathfrak{d} = b; \mathfrak{f} = bc; \mathfrak{g} = d$ and $\mathfrak{h} = abcd$ the three dependent elements marked by $\star$ under this setting $\mathfrak{a} = 1$, $\mathfrak{e} = \mathfrak{b}\mathfrak{d}+1 = ab+1$, and $\mathfrak{i} = \mathfrak{c}\mathfrak{g}+\mathfrak{f}\mathfrak{h}-\mathfrak{b}\mathfrak{f}\mathfrak{g}-\mathfrak{c}\mathfrak{d}\mathfrak{h}+\mathfrak{b}\mathfrak{c}\mathfrak{d}\mathfrak{g}+1 = cd+(bc)(abcd)-abcd-cb(abcd)+acbd+1 = cd+1$ are indeed the three matrix elements in $\cat^{3D}_P$ defined in Eq. \eqref{eq.CatMapPan3}.

\begin{table}[h]
\small
\centering
\caption{Denoting existing 3D cat maps with $\mathbf{C}{^{3D}_{W_{um}}}$}\label{tab.correspondence}
\begin{tabular}{r|llllll}
\hline\hline
& \multicolumn{6}{c}{\textbf{Existing 3D Cat Maps}}\\
\textbf{Para.} & ${\cat^{3D}_L}$& ${\cat^{xy}_{C}}$& ${\cat^{yz}_{C}}$& ${\cat^{zx}_{C}}$& ${\cat^{3D}_U}$& ${\cat^{3D}_P}$\\\hline
$u$ & 1 &1 &1 &1 &1 &1 \\
$m$ & 2 &1 &1 &1 &1 &2 \\\hline
$\mathfrak{a}$ & $\star$         &$\star$     &$\star$      &$\star$      &$\star$   &$\star$\\
$\mathfrak{b}$ & $0$       &0     &0      &$a_z$  &$a$ &$a$ \\
$\mathfrak{c}$ & $a$       &0     &$a_y$  &0      &0   &$c$ \\
$\mathfrak{d}$ & $bc$      &0     &0      &$b_z$  &$b$ &$b$ \\
$\mathfrak{e}$ & $\star$         &$\star$     &$\star$     &$\star$     &$\star$   &$\star$ \\
$\mathfrak{f}$ & $abc+c$  &$a_x$  &0      &0      &0   &$bc$ \\
$\mathfrak{g}$ & $bcd+b$  &0      &$b_y$  &0      &$c$ &$d$ \\
$\mathfrak{h}$ & $d$      &$b_x$  &0      &0      &$d$ &$abcd$ \\
$\mathfrak{i}$ & $\star$        &$\star$     &$\star$      &$\star$      &$\star$   &$\star$ \\\hline\hline
\end{tabular}
\end{table}

Meanwhile, the new proposed 3D cat maps can be extended to a more general case simply by multiplying 3D cat maps with interested configurations. In general, we can construct such a mixed 3D cat map $\cat^{3D}_{W_s}$ as shown in Eq. \eqref{eq.mixCat}, where $s_{um}$ is the indicator function for configuration $(u,m)$ defined in Eq. \eqref{eq.sum}. In this way, we are able to construct a more general 3D cat map with 36 new indicator parameters $\{s_{11},s_{12},s_{13},s_{14},s_{21},\cdots,s_{94}\}$.
\begin{equation}\label{eq.mixCat}
    \cat^{3D}_{W_s} = \prod\limits_{u=1}^9\prod\limits_{m=1}^4 \left(\cat^{3D}_{W_{um}}\right)^{s_{um}}
\end{equation}
\begin{equation}\label{eq.sum}
s_{um} = \left\{\begin{array}{l} 1 \textrm{, if } \cat^{3D}_{W_{um}} \textrm{ is interested}\\0 \textrm{, otherwise}\end{array}\right.
\end{equation}
\section{Simulation Results}
In many applications \cite{3DCat,CatLiu,CatPan}, a 3D cat map is used over a module $N$ as a finite state system shown in Eq. \eqref{eq.CatMapXYZ3D}, where the vector $[x_t,y_t,z_t]^T$ and $[x_{t+1}, y_{t+1} ,z_{t+1}]^T$ denote the discrete spatial coordinates at the time $t$ and $t+1$, respectively.
\begin{equation}\label{eq.CatMapXYZ3D}
        \left[\begin{array}{c}x_{t+1}\\y_{t+1}\\z_{t+1}\end{array}\right]= {\cat^{3D}} \left[\begin{array}{c}x_t\\y_t\\z_t\end{array}\right] \mod N
\end{equation}
The period length of this 3D cat map can be defined in Eq. \eqref{eq.period}, where $\mathbf{I}$ is the $3\times 3$ identity matrix.
\begin{equation}\label{eq.period}
    P_N(\cat^{3D}) = \arg\min\limits_{t\in \mathbb{Z}^+}\left\{\left(\cat^{3D}\right)^t \mod N = \mathbf{I}\right\}
\end{equation}
\begin{table}[h]
\centering
\small
\caption{Averaged Period Lengths of 3D Cat Maps}\label{tab.N10}
\begin{tabular}{r|rrrr|rrr}
  \hline\hline
\textbf{} & \multicolumn{7}{c}{\textbf{3D Cat Maps}}\\
$N$ & ${\cat^{3D}_L}$& ${\cat^{3D}_{C}}$& ${\cat^{3D}_U}$& ${\cat^{3D}_P}$& ${\cat^{3D}_{W_{11}}}$ &${\cat^{3D}_{W_{12}}}$ & ${\cat^{3D}_{W_{um}}}$\\\hline
10	&30.9	&53.7	&8.9	&20.7	&58.1	&59.2	&67.2\\
20	&50.7	&85.0	&14.4	&32.3	&88.5	&90.2	&99.2\\
30	&142.3	&320.9	&19.8	&55.4	&357.6	&362.2	&442.6\\
40	&81.3	&143.6	&24.4	&51.3	&156.5	&154.9	&176.7\\
50	&156.7	&274.2	&49.3	&109.3	&292.9	&293.0	&330.4\\
60	&217.4	&475.5	&30.0	&73.2	&492.1	&512.3	&589.0\\
70	&447.8	&961.5	&46.1	&244.5	&1045.7	&1079.4	&1228.3\\
80	&155.3	&276.1	&47.3	&92.2	&300.1	&301.2	&330.4\\
90	&315.2	&673.0	&46.8	&111.8	&764.8	&762.1	&929.5\\
100	&247.3	&420.3	&77.0	&164.9	&439.3	&439.4	&505.7\\
\hline\hline
\end{tabular}
\end{table}

Since any orbit in a finite state system is periodic, the system randomness can be largely reflected by averaged period length. And it is desired to have a 3D cat map with longer period lengths \cite{NagarajPeriod}. In regarding to the randomness of 3D cat maps, we perform the following comparisons on averaged period lengths using computer simulations. Specifically, we test six 3D cat maps ${\cat^{3D}_L}$, ${\cat^{3D}_{C}}$, ${\cat^{3D}_U}$, ${\cat^{3D}_P}$, ${\cat^{3D}_{W_{11}}}$, ${\cat^{3D}_{W_{12}}}$ and $\cat^{3D}_{W_{um}}$, and measure the period length of each cat map with a random set of parameters. Repeat this experiment 10,000 times and calculate the averaged period length for a 3D cat map under the module $N$ denoted as $\overline{P_N(\cat^{3D})}$ defined in Eq. \eqref{eq.period}.
\begin{equation}\label{eq.period}
    \overline{P_N(\cat^{3D})} = \sum\limits_{j=1}^{10000}P_N(\cat^{3D}_j)
\end{equation}
where $\cat^{3D}_j$ is the $j$th randomly generated $\cat^{3D}$ matrix. Simulation results of these averaged period lengths for $N = \{10,20,\cdots,100\}$ are given in Table \ref{tab.N10}. These results clearly indicate that the new proposed 3D cat maps ${\cat^{3D}_{W_{11}}}$ , ${\cat^{3D}_{W_{12}}}$, and ${\cat^{3D}_{W_{um}}}$ have much longer averaged period lengths than cat maps \cite{CatLian,3DCat,CatLiu,CatPan}.

\section{Conclusion}
In this letter, we have proposed a new family of 3D cat maps with eight parameters, two parameters controlling the cat map spatial configuration and the other six controlling the cat matrix elements. It incorporates the conventional 3D cat maps proposed by Lian \etal \cite{CatLian}, Chen \etal \cite{3DCat}, Liu \etal \cite{CatLiu} and Pan \etal \cite{CatPan} as special cases. It also outperforms these maps by providing more independent parameters and longer averaged period lengths. Both improvements on 3D cat maps are very meaningful for enhancing the security of chaotic cryptosystems and image encryption algorithms \cite{CatLian,3DCat,CatLiu,CatPan,kumar2010substitution,Fu2011,kanso2011novel,liu2011double}. A 3D cat map based system using new proposed 3D cat maps will have a larger key space to resist brute-force attacks and a longer averaged period length to resist statistical attacks. The presented framework is "universal" and allows extending to higher dimensional cat maps \cite{CatExt,Kwok2007}.
\bibliographystyle{IEEEtran}
\bibliography{report}
\end{document}